\documentclass[referee,pdflatex,sn-mathphys-num]{sn-jnl}
\usepackage{setspace}

\usepackage{graphicx}%
\usepackage{multirow}%
\usepackage{amsmath,amssymb,amsfonts}%
\usepackage{amsthm}%
\usepackage{mathrsfs}%
\usepackage[title]{appendix}%
\usepackage{xcolor}%
\usepackage{textcomp}%
\usepackage{manyfoot}%
\usepackage{booktabs}%
\usepackage{algorithm}%
\usepackage{algorithmicx}%
\usepackage{algpseudocode}%
\usepackage{listings}%

\theoremstyle{thmstyleone}%
\theoremstyle{thmstyletwo}%

\theoremstyle{thmstylethree}%

\newcommand{\snotes}[1]{\textcolor{black}{#1}}
\usepackage{color}
\usepackage{xspace}
\usepackage[normalem]{ulem}

\begin{document}

\title[Orbital House of Cards]{An Orbital House of Cards: Frequent \snotes{Satellite }Close Conjunctions}

\author*[1]{\fnm{Sarah} \sur{Thiele}}\email{sarah.thiele@princeton.edu}

\author[2]{\fnm{Skye R.} \sur{Heiland}}

\author[2]{\fnm{Aaron C.} \sur{Boley}}

\author[3]{\fnm{Samantha M.} \sur{Lawler}}

\affil*[1]{Department of Astrophysical Sciences, Princeton University, Princeton, NJ, 08540, USA}

\affil[2]{Department of Physics and Astronomy, University of British Columbia, Vancouver, BC, V6T 1Z1, Canada}

\affil[3]{Campion College and the Department of Physics, University of Regina, Regina, SK S4S 0A2, Canada}

\abstract{The number of objects in orbit is rapidly increasing, primarily driven by the launch of megaconstellations, an approach to satellite constellation design that involves large numbers of satellites paired with their rapid launch and disposal\snotes{, as well as the overall proliferation of satellite systems}. While satellites provide many benefits to society, their use comes with challenges, including the growth of space debris, collision \snotes{risks}, ground casualty risks, optical and radio-spectrum pollution, and the alteration of Earth's upper atmosphere through rocket emissions and reentry ablation. There is potential for current or planned actions in orbit to cause serious degradation of the orbital environment or lead to catastrophic outcomes, highlighting the urgent need to find better ways to quantify stress on the orbital environment. Here we propose a new metric, the CRASH Clock, that measures such stress in terms of the timescale for a possible catastrophic collision to occur if there are no satellite maneuvers or there is a severe loss in situational awareness. Our calculations show that the CRASH Clock is 5.5 days as of June 2025 \snotes{ and continues to decrease,} which suggests there is limited time to recover from a wide-spread disruptive event, such as a solar storm. This is in stark contrast to the pre-megaconstellation era: in 2018, the CRASH Clock was 164 days.}


\maketitle

\section{Introduction}\label{sec:intro}

The long-term sustainable use of satellites in Earth orbit requires an ongoing effort by all operators to limit the negative impacts of their actions. With this, there has long been a recognized need for identifying environmental targets and metrics. One example is the so-called 25-year rule (now being reduced to five years by some regulators), which sets an upper limit to the desired post-mission orbital lifetime of low Earth orbit (LEO) satellites in an effort to avoid producing debris from collisions \snotes{and to help free up orbital space after a mission ends} \cite{IADC2025,ESA2023}. A related example is the post-mission disposal (PMD) success rate, a measure of removing objects from orbit after their mission ends (see discussions regarding target PMD success rates in e.g. \cite{IADC2021})\footnote{Note that there are several definitions of PMD rates, applied in different contexts.}. This metric is often used to help parametrize long-term evolution models of Earth's satellite environment. The level of adherence by operators to PMD targets can strongly influence orbital debris evolution \citep{LIOU2005, Letizia2019}.

A more wide-reaching way to measure environmental stress on orbit is to identify an orbital carrying capacity, which broadly seeks to answer the question of how many things can be safely placed in orbit. The problem is that \snotes{we view }carrying capacity \snotes{as }not \snotes{being }an inherently well-defined metric because it may depend on tolerances for damage, how the satellites are maintained, and the types of objects in orbit. An example of a proposed measure for carrying capacity is the $\chi$ metric  (or ``$\chi$-capacity") \cite{DAmbrosio2024}, which is a ratio of the number of satellites in orbit at equilibrium (accounting for collisions) to the ideal (non-collisional) number. Another that calculates the maximal satellite population associated with a stable equilibrium debris population is the instantaneous Kessler capacity (IKC) \citep{Parker2025}. Others have been proposed, and we do not attempt to summarize them all.

Key environmental indicators (KEIs) offer a complementary approach by using metrics that are clearly defined and do not depend on specific use cases. Instead, they indicate the current orbital conditions and characterize stress (e.g., levels of orbital degradation and fragility) rather than capacity. Moreover, KEIs offer valuable tools for creating effective policies to address complex issues, such as the Sustainable Development Goal indicators \cite{SDG}. The measured PMD rate is an example of a KEI. Tracking how many satellites are visible above the horizon, averaged across the Earth as proposed by \cite{Lawrence2023} is a metric that might be interpreted as a KEI using our language. A symbiotic approach to this KEI framework is the concept of orbital ``health" indicators, proposed in \cite{Hugh2025}, which also includes the raw counts of artificial objects in space as a proposed indicator.

In some cases there is little distinction between the way carrying capacity is defined and how we use KEIs, such as when carrying capacity is used as a reference point to define consumption levels, creating a measurable proxy for orbital degradation. For example, \cite{Letizia2019} uses a globally integrated debris index as a measure of environmental capacity and its consumption, and propose using a maximum increase in this cumulative index compared to some reference epoch as a recommended ceiling.

The use of KEIs can further help to provide a suite of metrics for exploring when nonlinearities could occur or where there could be system-wide effects (such as across LEO). As an example, the concept of Kessler-Cour-Palais Syndrome (KCPS) \cite{Kessler1978} is made evident by KEIs (e.g., critical densities or collision area). KCPS is a condition in which the growth of collisional debris (and surface area) on orbit outpaces debris removal through atmospheric drag, causing a collisional runaway.  However, despite the use of the term ``runaway'', the initial phase of KCPS (which some argue we have already entered, see e.g. \cite{Kelvey2024}) is characterized by slow growth of debris, taking decades to centuries to develop. Such long timescales create a challenge for using the idea of KCPS alone as a policy tool.

While a KEI framework does not require the identification of a maximum stress, it does help to conceptualize a stress spectrum  -- where there is an acceptable range, an unknown range, and a clearly unacceptable range. This is helpful, as determining hard limits within a KEI framework is not a well-defined task, as discussed above. Indeed, for some KEIs, a clear threshold may not exist.

A KEI framework further provides a way to avoid shifting baseline syndrome (SBS) \cite{Masashi2018}. SBS describes the sociological and psychological phenomenon in which the acceptable baseline for environmental conditions degrades over time. It is therefore important to compare the current state of LEO not just to that of the era before megaconstellations,\footnote{\snotes{Here, the term ``megaconstellation'' refers to an approach to satellite systems that involves large numbers of satellites with a rapid replacement cycle. There is no clear number that defines a megaconstellation, but in usage refers to systems in the range between hundreds and tens of thousands of satellites. In this context, ``mega'' means big, with the same scientific usage as megalodon, megalith, megafauna, etc., and is not intended to mean 1 million. Using a clear term to define a change in the approach to satellite systems is of potential importance, as it could help to guard against shifting baselines, which is why it is used here instead of ``large'' constellations.}} but to the pristine orbital environment that existed  decades ago\footnote{Though outside the scope of this work, we note that this concept applies even more so to orbital light pollution and Dark and Quiet Sky protection \cite{DQSII}.}. Our tolerance for on-orbit risks should be calibrated with the very beginning of space exploration as its zero-point. Attempts should be made to counteract the bias of growing up with an already-littered orbit when setting space sustainability metric targets.

With this in mind, we introduce the Collision Realization And Significant Harm (CRASH) Clock, a KEI that evaluates the stress on the orbital environment. Many simulations predicting collision rates on orbit assume perfect collision avoidance for all maneuverable payloads during their operational lifetimes (e.g. Section 7.2 of \cite{ESA2025}, \cite{Letizia2023}). In contrast, the CRASH Clock uses the known distribution of resident space objects (RSOs: active and derelict satellites, debris, and rocket bodies) to determine how quickly we could expect a collision if all maneuvers were to suddenly stop or if there was a severe loss in situational awareness (such as due to a major solar storm or catastrophic software issue). It is a measure, in part, of the degree to which the orbital environment is a house of cards.

\section{Methods} \label{sec:methods}

\subsection{RSO Density Distribution}\label{sec:densities}

The rate of encounters (also referred to as `conjunctions' or `close approaches' in this work) or collisions on orbit depends heavily on the number density of objects and on the collisional area. We calculate the distribution of RSO number densities using the \snotes{Three Lines (3LE) orbital element set }of cataloged objects\footnote{All \snotes{element sets }were retrieved from \url{https://www.space-track.org/}.} for two chosen dates: 25 June 2025, and again for 1 January 2018 for a historical comparison. We divide LEO into a series of spherical shells with 1 km radial widths. Within each shell, the total number of RSOs is counted and divided by the shell volume. To capture the effects of eccentric orbits, including those that have apogees higher than LEO, an RSO's contribution to a given shell is weighted by the fraction of time per orbit that the RSO spends in that shell. 

Because objects are not randomly distributed in their orbital inclination, the resulting densities should be interpreted as averages, with some orbital configurations having significant density variations, such as the high densities that can occur near the maximum projected latitude excursions for inclined orbits. Our analytic encounter rate and CRASH Clock calculations presented below could potentially be improved by accounting for the actual distribution of inclinations. We leave this for future work, and do not think this changes the overall suitability of the CRASH Clock as a KEI.

\subsection{Analytic Close Encounter Rate}\label{sec:colrates}

Using the density distributions above, we can calculate expectations for close encounter rates among RSOs, as well as collisions under the assumption of no maneuvers. The satellite-satellite encounter/collision rate is approximated by \snotes{the statistical rate equation, akin to gas particle interactions (see e.g. \cite{Kessler1978} among others):}
\begin{equation}
    \Gamma_{\rm sat} = \frac{1}{2}\int_V n_{\rm sat}^2 A_{\rm col} \bar{v}_r \, dV,\label{eq:collRate}
\end{equation}
where $n_{\rm sat}$ is the number density of satellites, $A_{\rm col}$ is the encounter/collision cross-section of the satellites, $\bar{v}_r$ is the typical relative collision speed, and $dV$ is the spherical volume element. The factor of 1/2 is to ensure an encounter between two objects of the same population is not counted twice.
We also assume that the satellites are randomly distributed within a shell. The expected time between collisions is then $\tau_{\rm sat} = 1/\Gamma_{\rm sat}$.

This encounter/collision rate can naturally be extended to include combinations of various RSOs (e.g., debris-satellite, debris-debris), as well as non-trivial distributions of collisional area. Let the expected rate of close encounters less than some distance $d$ (i.e. $A_{\rm col}=\pi d^2$) within a spherical shell at altitude $h$ be given by $\Gamma_{d,h}$. The corresponding encounter time within each spherical shell is thus $\tau_{d,h}=1/\Gamma_{d,h}$. Furthermore, the time between close encounters less than distance $d$ for all of LEO can be found by summing over the rates, with $\tau_d=\big{(}\sum_h \Gamma_{d,h}\big{)}^{-1}$.

Assuming encounters are governed by Poisson statistics \snotes{\citep{SpaceDebris2006}}, the probability of seeing at least one encounter or collision by time $t$ for a given interaction with associated $\tau$ is $P = 1 - \exp\left( -t/\tau \right)$.

\subsection{Typical Relative Speed Calculation}

We now calculate the expected relative encounter/collision speed for Eq.~\ref{eq:collRate}. Assume that the orbits of objects in a shell are randomly oriented and circular. Under this assumption, the encounter velocity between any two RSOs is  
\begin{equation}
\vec{v}_r = v_o\left( \sin\theta \hat{x} + (1-\cos\theta) \hat{y}\right),
\end{equation}
where $v_o$ is the circular orbital speed for the shell, $v_o(h)=\sqrt{GM_\oplus/(R_\oplus+h)}$ for altitude $h$. We have oriented our frame of reference such that $\hat{y}$ is one satellite's direction of motion, with $\theta$ representing the angle between the two satellites' directions of motion.  The relative speed of the encounter is then
\begin{eqnarray}
    v_{r} &=&  v_o \left( \sin^2\theta + (1 - 2 \cos\theta + \cos^2\theta)    \right)^{1/2}\\
    & = & \sqrt{2} v_o (1-\cos\theta)^{1/2}.
\end{eqnarray}

\begin{figure}
    \centering
    \includegraphics[width=0.5\linewidth]{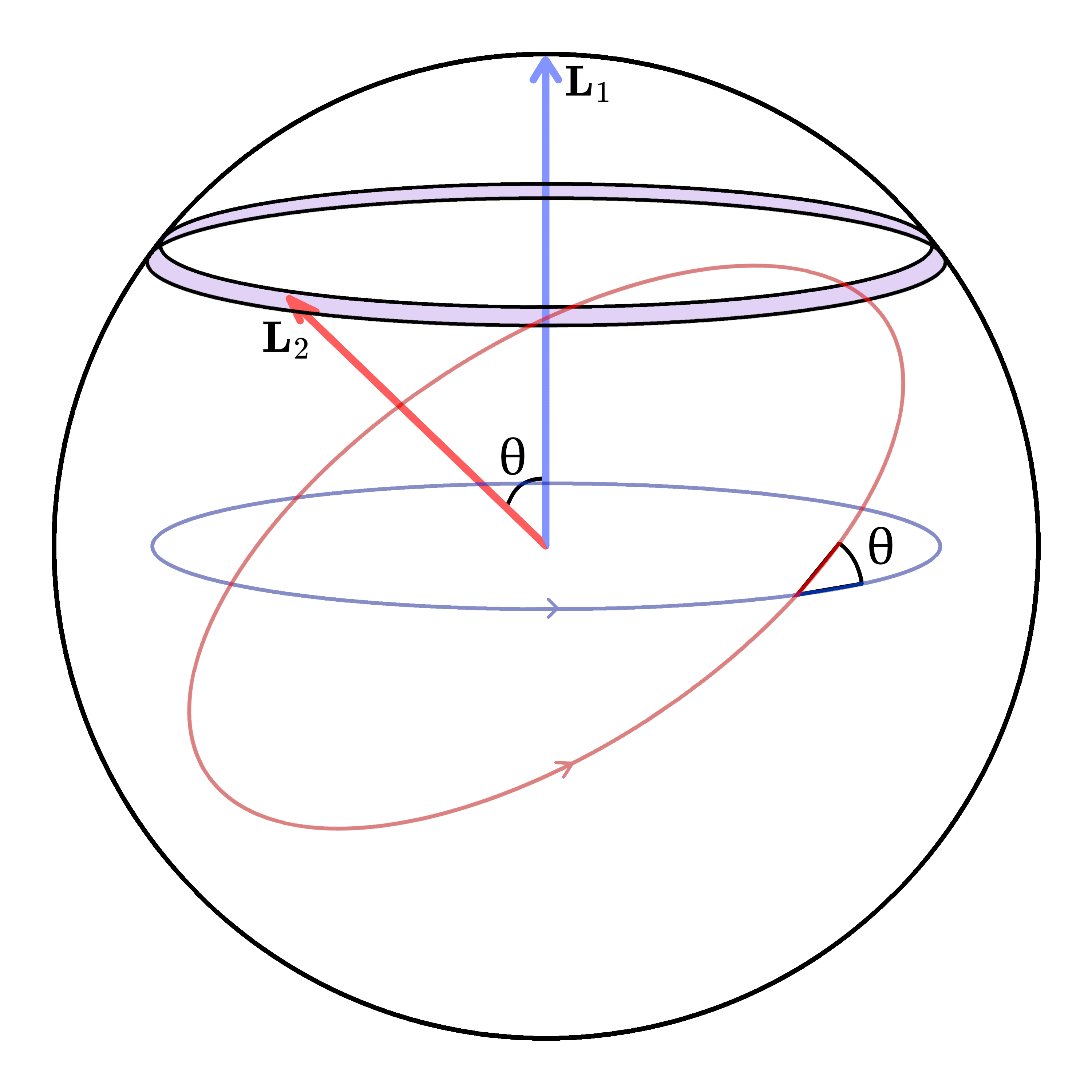}
    \caption{\snotes{The angle between two satellites' direction of motion $\theta$ is the angle between their orbit normal (or angular momentum) vectors. The probability of a randomly drawn vector forming an angle between $\theta$ and $\theta + d\theta$ is given by the area subtended by that range of angles (shown in purple), $2\pi \sin\theta \, d\theta$, divided by the full area of the unit sphere, $4\pi$. Hence, $p(\theta) = \frac12 \sin\theta$ on $[0, \pi]$.}}
    \label{fig:angleDist}
\end{figure}

The average is calculated by integrating over all possible orientations. We assume that the angular momentum vectors are randomly \snotes{ and isotropically} distributed over a sphere (this is slightly different from assuming that the angle $\theta$ is \snotes{uniformly }distributed\snotes{, see Fig. \ref{fig:angleDist}}). The resulting integration is thus
\begin{eqnarray}
    \bar{v}_r & = & \frac{\sqrt{2}}{2} v_o \int_0^\pi \left( 1-\cos\theta\right)^{1/2}\sin\theta d\theta\\
    \bar{v}_r & = & \frac{4}{3} v_o.
\end{eqnarray}
At an altitude of 550 km, for example, the resulting typical relative speed between RSOs is about 10 km/s.

\subsection{The CRASH Clock}\label{sec:clock}

Taking the encounter rate estimates above one step further, we can approximate the time for close approaches that could give rise to collisions on orbit. We propose this approach as a measure of the stress in orbital space: the CRASH Clock. In other words, the CRASH Clock is the expectation time for a possible collision to occur if all satellite maneuvers were to suddenly stop and satellite orientations could not be controlled. For this calculation, we must make some assumptions about collision cross sections.

We use our previously calculated RSO number density distributions as described in Section \ref{sec:densities}, but keep the densities separated into categories (i.e. the total number density in a given shell is the sum of $n_{\rm debris}$, $n_{\rm sat}$, etc.). Each interaction combination is then assigned a ``collision'' cross section (described in the next section).

The total collision rate for a given altitude is
\begin{equation}
\Gamma_h = \bar{v}_rV_{h}\sum_{i,j\leq i} \left(1-\frac{1}{2}\delta_{ij}\right)n_in_jA_{ij}^{\rm col}
\end{equation}\label{eq:collRate_full}
for object types $i$ and $j$, shell volume $V_{h}$ and average relative speed $\bar{v}_r=\bar{v}_r(h)$. Terms with $i=j$ are halved as in Eq.~\ref{eq:collRate}. The corresponding collision time is found by summing the rates over all of LEO.

\subsection{Collision Cross Sections}\label{sec:cross-section}

All calculations in this work that utilize Eq. \ref{eq:collRate} require an assumed collision cross section. For a given close encounter distance $d$, we assume a cross section of $A_{\rm col}=\pi d^2$. However, for possible collisions among LEO RSOs, we must instead assign collision cross sections to each interaction combination, depending on the object type. In our base results (Section~\ref{sec:results}), we assume that the collision cross sections for satellite-satellite, rocket body-satellite, rocket body-rocket body, debris-satellite, debris-rocket body, debris-debris are 300, 300, 300, 79, 79, 0.03 $\rm m^2$, respectively. These collisional areas, which are not the same as the individual cross section of each object, correspond to a close approach distance of 10 m between two satellites, satellite-rocket body, or two rocket bodies; 5 m between a satellite or rocket body and debris; and 10 cm between two pieces of debris. We thus refer to the CRASH Clock calculated using this choice of corresponding cross sections as the ``10-5-10 Clock''. While cross sections will need to be reevaluated as conditions and information change, we use the 10-5-10 Clock because it captures close approach distances of serious concern, where a collision is possible but not guaranteed. 

\begin{figure}
    \centering
    \includegraphics[width=0.8\linewidth]{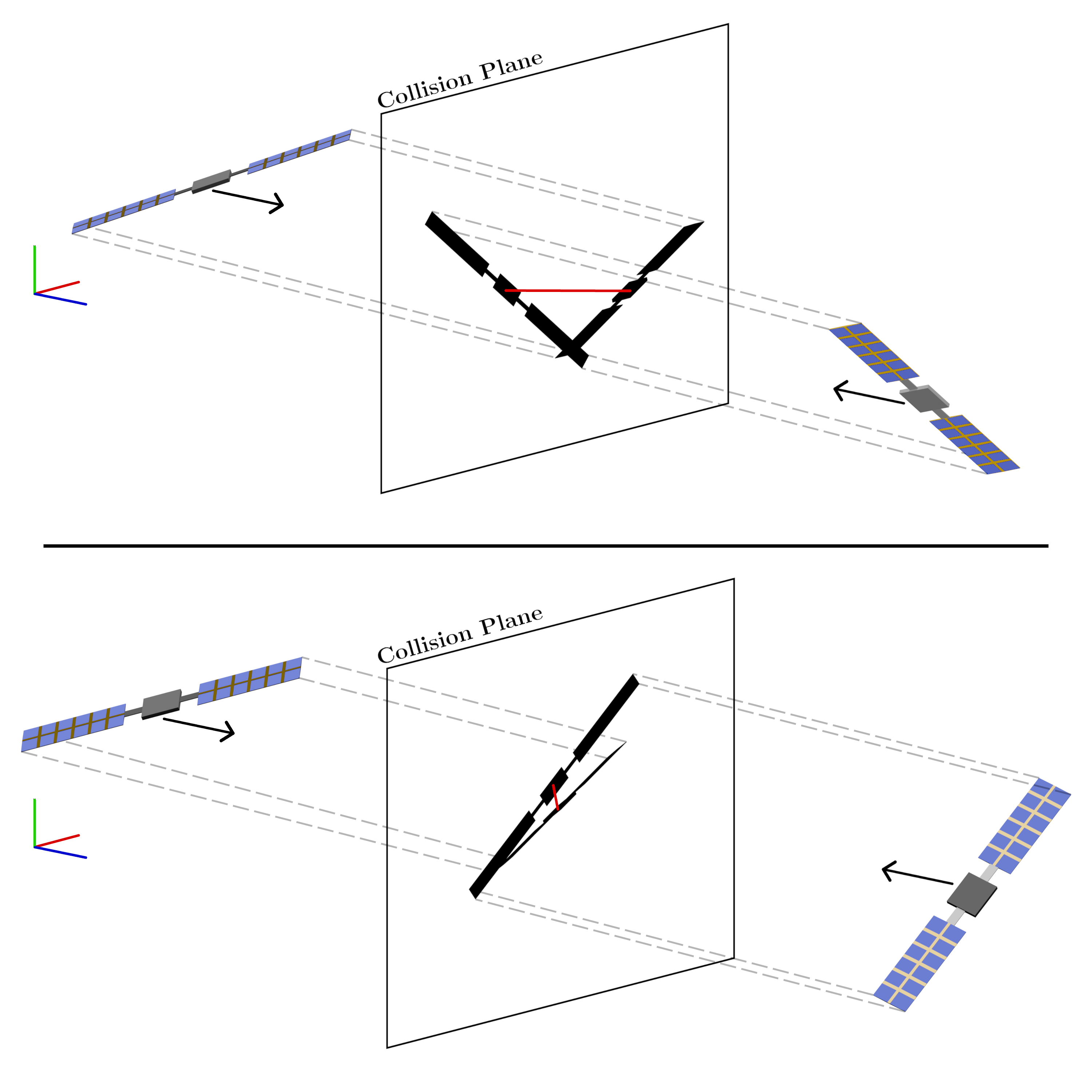}
    \caption{\snotes{Two examples of randomly oriented satellites with their directions of motion shown, projected into the collision plane between them. The projected center-to-center distance, which can vary considerably, is marked in red. This distance depends on the angle between the satellite orientations, and is further modified by the tilt of the satellite (that is how edge-on versus face-on the solar panels are). These orientations are for illustrative purposes, and represent neither the minimum or maximum approaches.}}
    \label{fig:collPlane}
\end{figure}

We do not use the average area of a satellite for our base calculations because doing so could underestimate the collision potential, possibly severely, due to the importance of the orientation of objects during a close conjunction. \snotes{A demonstration of different encounter orientations is shown in Fig. \ref{fig:collPlane}.} Take the Starlink V2 mini as an example, which has approximate dimensions of 0.3 m $\times$ 4.1 m $\times$ 29 m. The average surface area is approximately $43\rm~m^2$. If the collision cross section between two objects is defined as $A_{\rm col}=(A_1^{1/2}+A_2^{1/2})^2$, as is commonly done, the estimated collision cross section would be $61\rm m^2$. However, for example, two randomly oriented thin rods of full span $L$ have a typical collision distance (between rod centers) of approximately $0.3 L$. Thus, for the Starlink V2 minis, assuming random orientations, the effective collision cross section approaches the $300\rm~m^2$ value we use --  the average area of the satellite vastly underestimates the collision potential in this case.

Nevertheless, we are mindful that not all objects will have such geometries, and as such, it is important to explore different Clocks. Based on the ESA environmental report \citep{ESA2025} of average megaconstellation satellite surface areas, a rough estimate of collisional encounter distances is 4.8 m (i.e., a collision cross section of about $72\rm~m^2$). Thus, if we take close approaches of 4.8 m, 2.4 m, and 10 cm (in the same way as above) to lead to collisions, we could calculate a 4.8-2.4-10 Clock.

In short, the Clock value is dependent on assumptions of the minimum tolerable close approaches,  but the trends highlighted by various Clocks are ultimately the same.

\subsection{N-body Conjunction Simulation} \label{sec:nbody}

We verify our analytic model against direct N-body conjunction simulations. Written in Python, the simulation code {\tt SatEvol}\footnote{The code is open-source and can be found at \url{https://github.com/norabolig/conjunctionSim}.} propagates orbits using Keplerian orbital elements, and includes nodal and apsidal precession due to Earth's $J_2$ gravitational moment. Whenever objects are within a defined threshold distance of each other, for example 1 km, those pairs are tracked until their separation is once again greater than the threshold. The \snotes{orbital element set} for a given date \snotes{is} used for the initial conditions, consistent with those used for producing the density distributions. All RSOs are propagated to a common starting epoch using the Python implementation of the Simplified Gravitational Perturbations model (SGP-4) \cite{Vallado2006}. Minimum distances are determined using a $k-$dimensional tree search \citep{Maneewongvatana1999}. 

Postprocessing of simulation outputs is then used to identify each conjunction, along with its minimum encounter distance. We track objects according to their catalog type, e.g., satellites, debris, and discarded rocket bodies, as well as specifically Starlink satellites due to their abundance. At this time, the code does not include additional perturbative effects (e.g. from the Sun and Moon, tesseral harmonics, radiation pressure, or atmospheric drag). 

To capture these extremely fast close approaches, we use a time step of 0.05 seconds for the 1 km conjunction sims, and a step of 0.001 seconds for the 100 m conjunction sims, both presented below.

\section{Results} \label{sec:results}
\begin{figure} \centering
\includegraphics[width=0.45\textwidth]{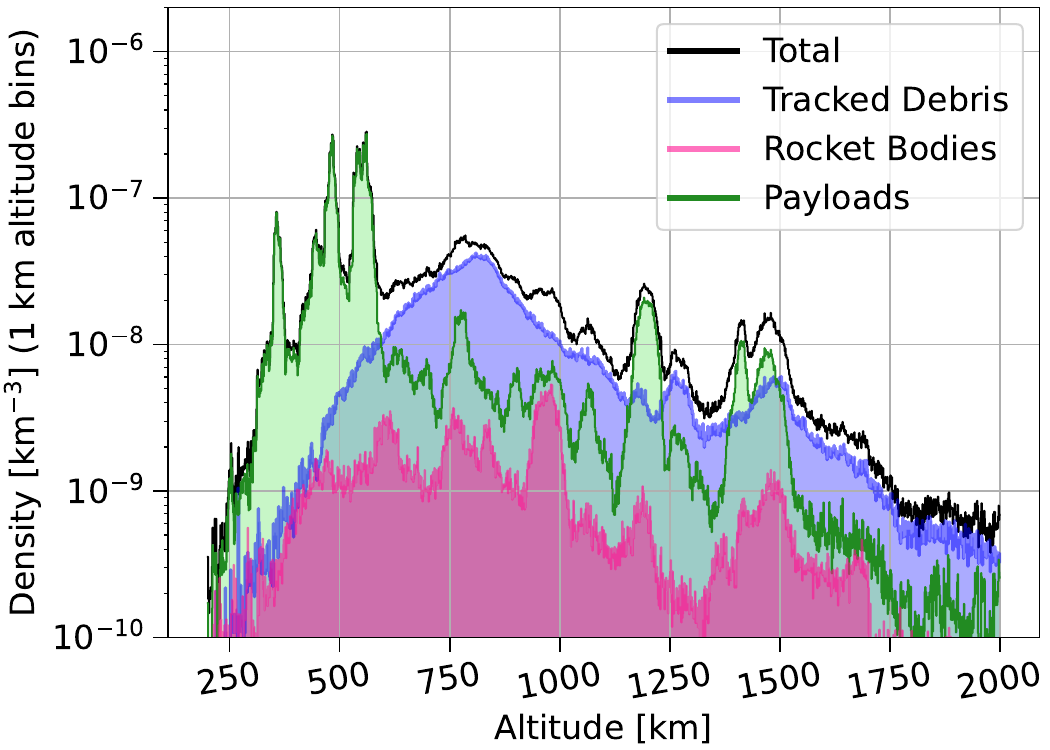}
\includegraphics[width=0.385\textwidth,trim={73 0 0 0},clip]{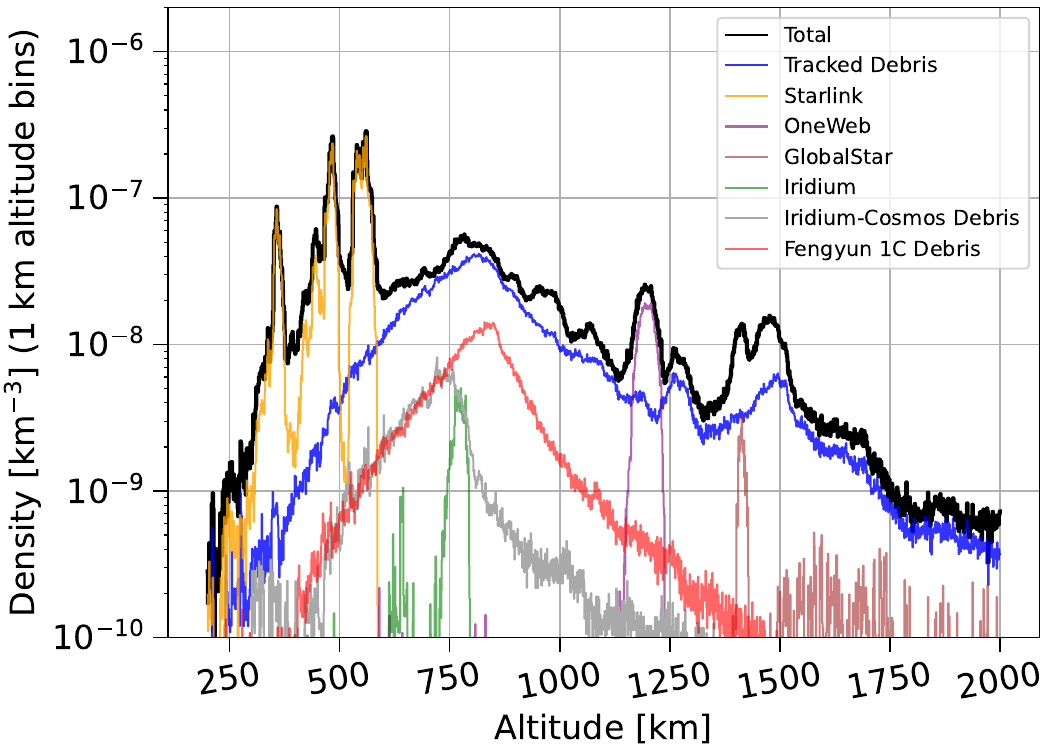}
\caption{Orbit-averaged volume density distribution of RSOs by classification (left) and origin (right) as of 25 June 2025. Left: `Payload' refers to active and defunct satellites, `Tracked Debris' only those debris pieces that are reliably tracked in the debris catalog, excluding abandoned rocket bodies and defunct satellites. Right: Densities of select active constellations and megaconstellations, as well as tracked debris, debris from the 2009 Iridium-Cosmos collision, and debris from the 2007 Fengyun 1C ASAT test.
}\label{fig:density}
\end{figure}

\subsection{Analytic Encounter Rate Results}\label{sec:analytic_results}
We start by exploring the current number density distribution for different types of RSOs, shown in Fig.~\ref{fig:density}, based on the orbital information in the satellite catalog for 25 June 2025, averaged over spherical shells. As discussed in Section \ref{sec:densities}, the resulting densities should be interpreted as averages, with the understanding that there can be substantial variation in local densities.

Starlink satellite shells exhibit the highest densities on orbit, reaching over an order of magnitude higher than the tracked space debris peak at approximately 800 km altitude, which has large contributions from the 2007 Chinese anti-satellite weapon test (Fengyun 1C) and the 2009 Iridium 33-Cosmos 2251 collision. 

To demonstrate the consequences of Fig.~\ref{fig:density}, we can integrate Eq.~\ref{eq:collRate} over a limited volume. As an illustrative example, let us integrate over the peak in the density distribution centered on 550 km altitude, assuming a uniform spherical shell. We approximate the shell as being 30 km thick from the peak's width, and assume a possible collision cross section of $A_{\rm col} \approx 300~\rm m^2$, corresponding to a close approach of 10 meters. As discussed above, while such a close approach does not guarantee a collision, depending on the orientation of the satellites involved, it could lead to one. For simplicity, we use the term ``collision" moving forward. Indeed, as noted in Section \ref{sec:cross-section}, the average collision distance for randomly oriented thin rods of full length $L$ is approximately $0.3 L$. Satellites that are thin but have 30 m spans can have very large effective cross sections, should there be a loss of control of their orientations.  Again, from a safety point of view, using an average physical area for the satellite would underestimate the collision cross section, possibly severely.

As also discussed in Section \ref{sec:cross-section}, we use $\bar{v}_r\approx 10~\rm km~s^{-1}$. The density appears to peak at $n_{\rm sat}\approx 3\times 10^{-7}~\rm km^{-3}$, but upon closer inspection we find that the density finely oscillates around $n_{\rm sat}\approx 2\times 10^{-7}~\rm km^{-3}$, so we use this latter value as the average density in the shell. 

For the given altitude, the single shell collision rate is $\Gamma_{\rm ss}\approx 2.2\times10^{-6}~\rm s^{-1}$ or about $0.09\rm~d^{-1}$, with which we find $\tau_{\rm ss} = 1/\Gamma_{\rm ss} \approx 11~\rm d$. This means that the timescale for a 50\% chance of one or more potential collisions is about $t\approx 7.6\rm~d$ in that shell alone. This result, while approximate, emphasizes that the collision probability for satellites on orbit is substantial without active management. 

Although spherical shells around Earth represent extremely large volumes and the instantaneous volume occupied by satellites is small, LEO satellites orbit the Earth in approximately 90 minutes depending on the altitude. As a result, these satellites quickly explore their mutual interaction possibilities.
Collision avoidance maneuvers and station-keeping (i.e. maneuvers to maintain desired altitudes, separations and orbital phasing), are essential in dense satellite shells and successful active management of constellations seems to be the only reason why there has not been a recent major satellite-satellite collision as orbital densities continue to increase. We will revisit the importance of station-keeping further below.

Now we extend our calculation to include all RSO interactions throughout all of LEO (see methods in Section \ref{sec:colrates}). Let the expected rate of close encounters less than 1 km in distance (i.e. $A=\pi~\rm km^2$) within spherical shells of altitude \textit{h}  be given by  $\Gamma_{1,h}$. We choose a conjunction distance of 1 km because it roughly corresponds to the distance at which a collision avoidance maneuver is executed. In practice, such maneuvers do not depend on a fixed distance; rather, they occur when the probability of a collision exceeds a risk tolerance threshold (e.g. \cite{SpaceX2026}), which varies between operators. That probability is further dependent on the uncertainties of the orbits in question. We avoid these complexities by using the 1 km approach distance as an imperfect proxy. 

\begin{figure}[ht]\centering
\includegraphics[width=\textwidth]{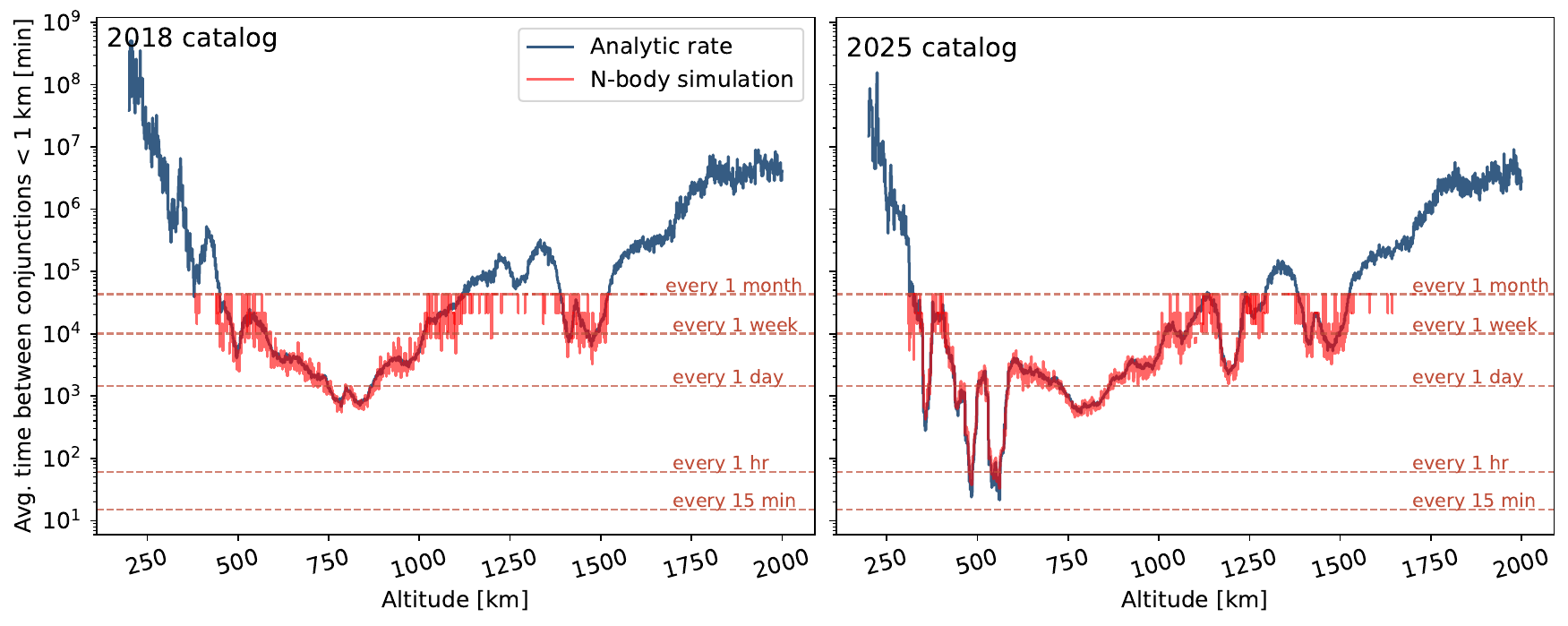}
\caption{Average time between conjunctions $< 1 \, {\rm km}$ as a function of altitude for the 2018 (left) and 2025 (right) RSO populations. The dips around $500 \, {\rm km}$ correspond to the peaks in RSO density at the same altitude presented in Fig.~\ref{fig:density} due to Starlink. Our analytic calculation is shown in dark blue, with the simulations overlaid in red. The simulation duration was one month.} 
\label{fig:closeapproach}
\end{figure}

With this in mind, the corresponding time between encounters within each spherical shell is $\tau_{1,h} = 1/\Gamma_{1,h}$ (see Section \ref{sec:colrates}). Fig.~\ref{fig:closeapproach} shows the analytic result using the blue curve, based on the density distribution from Fig.~\ref{fig:density}. The results of an n-body simulation is also shown in this figure, discussed further below.  In the densest part of Starlink's 550~km orbital shell, we expect close approaches ($<1~\rm{km}$) every 22 minutes in that shell alone.

The time between $< 1$ km close encounters for all of LEO is found by summing over the rates, with $\tau_1=\big{(}\sum_h \Gamma_{1,h}\big{)}^{-1}$. For all objects, Eq.~\ref{eq:collRate} gives $\tau_1=36$ seconds. This is dominated by encounters involving at least one satellite (i.e. Sat-RSO encounters), with Starlink (the primary operator in orbit) satellites making up the majority of the satellite population. For Sat-RSO encounters, $\tau_1 \approx 41~\rm{s}$, while for Starlink-RSO, $\tau_1 \approx 47~\rm{s}$.

According to the most recent SpaceX biannual report, Starlink satellites made 148,696 collision avoidance maneuvers in the period between 1 June 2025 and 30 November 2025 \citep{SpaceX2026}, averaging to 36 maneuvers per satellite per year, or one collision avoidance maneuver every 1.8 minutes across the whole megaconstellation.  Keeping in mind that our encounter distance does not directly translate to a collision probability threshold for maneuvers, nor do we take into account the effects of station-keeping, we interpret the Starlink maneuver time to indicate that our simulation results and analytic calculations (1.1 min and 47 s, respectively, see Methods and following discussion regarding our simulations) are reasonable for a no-maneuver situation.

\subsection{The CRASH Clock Results}\label{sec:clock_results}

For the June 2025 catalog, the 10-5-10 CRASH Clock is $\tau_{\rm col}=5.5\rm~d$. For comparison, the no-maneuver collision time for Starlink-RSO collisions is about $6\rm~d$. Said differently, if all types of maneuvers were to suddenly stop (and no orientation control), and with our assumptions for collision cross-sections, within 24 hours there is a 17\% chance of a close approach that could lead to a collision between two cataloged RSOs, with a 15\% chance that the interaction would involve a Starlink satellite, assuming a Poisson process. Such collisions would be catastrophic, causing a major debris-generating event with high likelihood of secondary and tertiary collisions due to high orbital densities and local collision areas. We repeat this calculation using the LEO \snotes{element set }catalog from 1 January 2018, providing a reference value from prior to the megaconstellation era. In stark contrast, the 10-5-10 CRASH Clock was 164 days, corresponding to less than a 1\% chance of a collision happening between two RSOs within 24 hours under no-maneuver conditions.

As discussed in Section \ref{sec:cross-section}, different cross sections could be assumed (and with this, different risk tolerances). As an example, using a 4.8-2.4-10 Clock gives a collision time of 23 d for 2025 and 701 d for 2018. Again, however, we note that these smaller cross sections may underestimate the potential of having a collision.

\subsection{Comparison to N-body Simulations}\label{sec:nbody_results}

The fidelity of our analytic calculations is compared to direct N-body simulations (see Section \ref{sec:nbody}). Simulations were initialized using \snotes{orbital element sets} for our two different epochs: 1 January 2018, and 25 June 2025. The red curves in Fig.~\ref{fig:closeapproach} show the average time between conjunctions less than $1 \, {\rm km}$ as a function of altitude, which shows very good agreement with the analytical estimate. We emphasize that while the analytic simulations assume random orbital configurations within each shell, the simulations use the actual orbits. Thus, at least for counting close conjunctions in a statistical sense, our assumption of a random distribution does not appear to be a limitation of the analytic approach. The left-hand panel of Fig.~\ref{fig:closeapproach} shows the 2018 distribution, the right-hand shows 2025. We see that the minimum time between conjunctions has decreased by about two orders of magnitude between 2018 and 2025 for a no-manoeuvre situation. 

To better compare the analytic and simulation methods for even closer approaches and possible collisions, additional simulation runs were carried out using a much smaller time-step (see Section \ref{sec:nbody}) and a conjunction distance threshold of $100 \, {\rm m}$. For the 25 June 2025 catalog, we see a $100 \, {\rm m}$ conjunction frequency of approximately one encounter every 42 minutes, within a factor of two of the analytic estimate (which expects approximately one encounter every 60 minutes for distances less than $100 \, {\rm m}$).  For this particular simulation initialization, a conjunction less than 30 meters  occurs between debris and a satellite in the first 3 hours. While this is not inherently a collision, it is an extreme encounter. We note that there is some difficult-to-characterize uncertainty in these numbers because we are using only a single simulation initialization.  We have summarized all of our results for both \snotes{orbital element }catalog epochs in Table \ref{tab:clock}.

Both the simulations and the analytic calculations are intended to represent encounter statistics for a snapshot in time, akin to a steady state system. That is, we make our calculations using only a chosen \snotes{orbital element }catalog epoch and do not incorporate evolution of the objects on orbit through launches, active debris removal, or PMD. We note that our simulations also do not incorporate atmospheric drag, which would serve to clear out debris at lower altitudes, decrease satellite altitudes, and potentially increase the rate at which satellite orbits would randomize without station-keeping efforts in place. 

The distribution of densities, and thus encounter/collision rates, also evolve as objects are launched into orbit and satellite configurations change: we ran additional simulations (not shown here) for an October 2024 catalog, and saw that the morphology of the encounter distributions were different than that of 2025 (though the shortest encounter interval stayed about the same). 

\begin{table}
    \centering
    \caption{Results for our analytic and simulated close encounter times. We show results for both our 2018 and 2025 epochs, for encounter distances $d$ within $1~\rm{km}$ and $100~\rm{m}$, as well as our CRASH Clock value for each epoch. Our simulations and analytic calculations agree to within a factor of two.}
    \label{tab:clock}
    \begin{tabular}{lcccc}
        \toprule
         & 2018, analytic & 2018, sim & 2025, analytic & 2025, sim \\
        \toprule
        $d < 1$ km, all LEO & 3.9 min & 3.5 min & 36 s & 44 s \\
        $d < 1$ km, Sat--RSO & 12.4 min & 11.0 min & 41 s & 53 s \\
        $d < 100$ m, all LEO &  6.6 hr & 5.2 hr & 60 min & 42 min \\
        $d < 100$ m, Sat--RSO &  20.7 hr & 10.5 hr & 69 min & 56 min \\ 
        \hline
        CRASH Clock & 164 d & & 5.5 d & \\
        \bottomrule
    \end{tabular}
\end{table}

\section{Discussion} \label{sec:discussion}

 Simulations have shown that altitudes above 600-800~km in LEO are already above the unstable threshold for long-term runaway debris growth, i.e., KCPS \citep{LiouJohnson2008,kessler2025}. Strikingly, given the density and surface area, the main Starlink shell (about 550~km altitude) is also within the runaway threshold \citep{kessler2025}, meaning a single collision could have substantial longer-term consequences. While collisional cascades can take decades to centuries to develop, a single collision could create substantial stress on the orbital environment immediately, even if it does not lead to a runaway \citep{Thiele2022,Boley2024}.

We suggest that the CRASH Clock presents a complementary and potentially better way to characterize orbital stress for both short and long-term consequences \footnote{The CRASH Clock is hosted and periodically re-evaluated at the public website \url{https://outerspaceinstitute.ca/crashclock}.}. This KEI, calculated using the methods above, provides an immediate assessment, as well as historical record, of the stress on the orbital environment. It further provides a measure of the typical time between system-wide loss of control in LEO and a possible catastrophic collision. 

We find that our simulations and analytic calculation yield consistent results to within a factor of two. For the purpose of determining the CRASH Clock value, we recommend using the analytic method, even though it assumes randomized orbits. There are several reasons for this. First, the analytic approach is an accessible calculation that only requires \snotes{an orbital element set }catalog.  Second, it avoids differences in the results that could arise from the choice of numerical methods. Third, simulation techniques could suffer from substantial statistical variation, potentially requiring many simulation initializations. 

Station-keeping and constellation design deserve a special mention. Satellite constellations can be designed to minimize conflicting interactions among their satellites, with station-keeping being done to maintain optimal configurations \cite[see e.g.][]{sk1,sk2,sk3}. We can thus contemplate situations in which we have ideal constellation designs. In such cases, the CRASH Clock value could in practice be modified to take into account the ``phase mixing" time of the constellation in question; by phase mixing time, we mean the theoretical time it would take for the constellation's orbits to randomize in phase. However, we emphasize again that our analytics and simulations are in reasonable agreement, the latter of which takes into account a snapshot of actual constellation configurations. For this reason, we recommend against removing constellations with idealized designs when determining the CRASH Clock value. 

\begin{figure}
    \centering
    \includegraphics[width=0.8\linewidth]{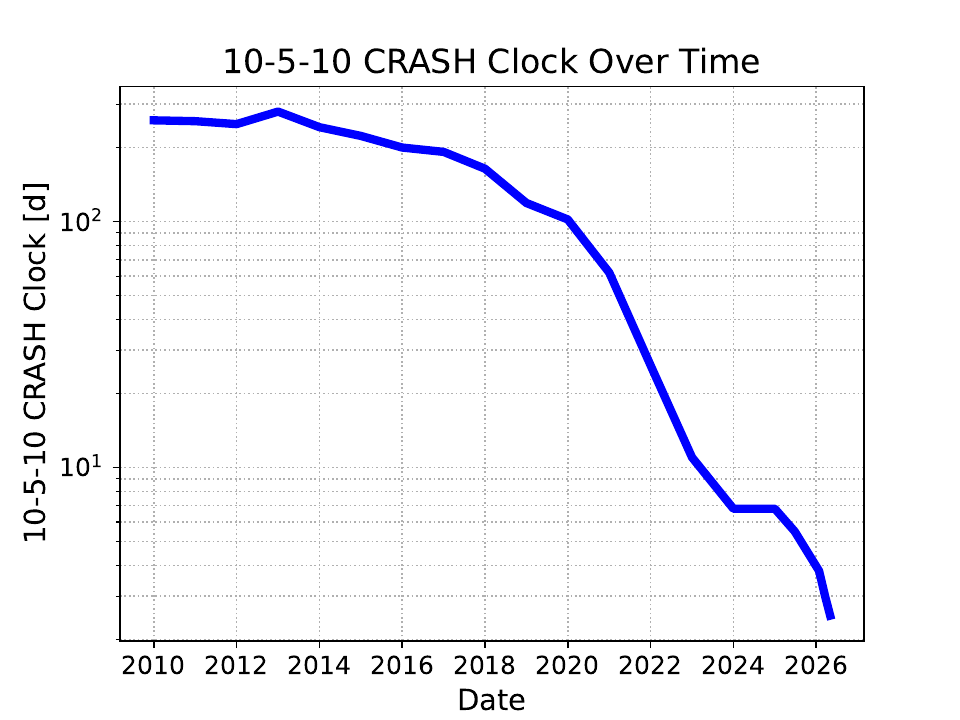}
    \caption{\snotes{The 10-5-10 CRASH Clock as a function of time. The Clock dropped below 100 days in 2020 and has continued to decline. The rapid drop in the Clock value is due to the increased deployment of satellites. As a result, also around 2020, the system transitions from satellite-debris to satellite-satellite dominated in the Clock framework. As of May 2026, the value is below 3 days.}}
    \label{fig:crashclock_time}
\end{figure}

As of June 2025, the 10-5-10 CRASH Clock in LEO is 5.5 days. Yet, the probabilistic nature of collisions means they could happen sooner than any given CRASH Clock value (again in the absence of maneuvers). Recall that our example numerical simulations showed a debris-satellite conjunction of less than 30 meters in the first 3 hours of simulation time. \snotes{Moreover, we calculated the CRASH Clock as a function of time between 2010 and May 2026, shown in Figure \ref{fig:crashclock_time}. The CRASH Clock continues to decrease, with the 4 May 2026 value now at 2.5 days.}

We emphasize that the CRASH Clock does not measure the onset of KCPS, nor should it be interpreted as indicating a runaway condition. However, it does measure the degree to which we are reliant on errorless operations. In the short term, a major collision is more akin to the Exxon Valdez oil spill disaster \cite{exxon} than a Hollywood-style immediate end of operations in orbit.  Indeed, satellite operations could continue after a major collision, but would have different operating parameters, including a higher risk of collision damage. 

A CRASH Clock value of less than a week is already a reason for concern, \snotes{let alone a few days}, as major solar storms, such as the May 2024 Gannon storm, can have lingering impacts for the satellite population. For example, in the three days of this storm, nearly half of all active LEO satellites maneuvered due to increased atmospheric drag to maintain their orbits, and the unpredictable drag in addition to bulk maneuvering made collision assessment during and after the storm very difficult \citep{Parker2024,Berger}. Additionally, in such conditions positional uncertainties can easily become as high as several km \citep{Parker2024_2}, making collision avoidance maneuvers extremely uncertain. Moreover, while the May 2024 storm was the strongest geomagnetic storm in decades, The Great Geomagnetic Storm of 1859 was at least twice as intense \citep{DSTDATA,CLIVER2013}. That event was characterized by two strong storms within a few days of each other, with the latter commonly known as the Carrington Event (September 1859). The storm peaks lasted for several hours, while the storm durations were for a day and several days, respectively \citep{GREEN2006}. 

The number of collision avoidance maneuvers made by Starlink has historically been doubling every six months \citep{Pultarova}, although this seems to be the result of several factors, including lowering their risk thresholds (i.e., increasing safety practices). Each maneuver can create uncertainty in the estimated satellite positions, with one study even finding inaccuracies immediately after the maneuver of up to 40 km \citep{Pultarova2}. As the number of required maneuvers continues to increase, temporary lapses in collision avoidance capabilities, whether that be from inaccurate orbital determination or even a small miscommunication between operators in maneuver decision-making, will become increasingly catastrophic in their potential consequences. Indeed, in 2019 an ESA satellite was forced to maneuver out of the way of a Starlink satellite when a bug in SpaceX's alert system prevented them from seeing an increased collision probability \citep{boyle}. Before modern space traffic management policies, insufficient maneuver plan information between operators was the main cause that led to the Iridium-Cosmos collision of 2009 \citep{shepperd}.

The Guidelines for the Long-Term Sustainability of Outer Space Activities of the United Nations Committee on the Peaceful Uses of Outer Space took the important step of identifying Earth's orbit as a finite resource \cite{UNOOSA}. Placing a satellite into orbit is resource consumption; debris, abandoned rocket bodies, and derelict spacecraft all lock up resources without any benefits. However, numbers of objects alone provide insufficient information. Collisional surface area and timely updates to published orbits are also needed. The CRASH Clock is, in part, a measure of the consumption of Earth's orbital space and the degree to which operations there are being done sustainably. Increases in either orbital density or collisional cross section decreases the collision time on orbit, and reduces the margin of error for safe operations.

All the above discussion contextualizes how the CRASH Clock can be used. It is not a strict limit - rather, it measures risk along a spectrum, with short CRASH Clock times representing a dangerous condition, moderate times representing a caution region, and long times indicating a healthy operational orbital environment. There is some degree of subjectivity in the boundaries of these ``danger", ``caution", and ``safe" regions, but this makes the CRASH Clock adaptable without changing the underlying definition. As an illustrative example, and recalling our calculations above, \snotes{the June 2025 }10-5-10 CRASH Clock value of $\tau_{\rm col}=5.5~\rm d$ corresponds to a 17\% probability of one or more potential collisions during a 24 hour period of no-maneuver conditions. We consider this to be well within the ``caution'' region. We could further (and arbitrarily) define the ``danger'' region to be where the CRASH Clock value implies a 50\% chance of at least one potential collision within 24 hours. To stay in the ``caution'' region and below this ``danger'' threshold, the CRASH Clock value would therefore need to be longer than 1.4 d. \snotes{The May 2026 value $\tau_{\rm col}=2.5~\rm d$ shows that we are rapidly approaching this situation.}

We could adopt smaller cross sections for the CRASH Clock, indicating when collisions are not just possible but probable.  The 4.8m-2.4m-10cm Clock (see Methods) is one example, with a value of about 23 d. From a policy standpoint, however, there needs to be discussion concerning whether the Clock threshold should indicate possible or probable collisions. We have highlighted possible collisions not to be alarmists, but to contextualize the stress and demands on space safety.

Collision risk is not the only issue. We are already experiencing disruption of astronomy \citep{Lawrence2022}, pollution in the upper atmosphere from increasingly frequent satellite ablation \citep{Murphy2023}, and increased ground casualty risks \citep{Wright2025}. By these safety and pollution metrics, it is clear we have already placed substantial stress on LEO, and changes to our approach are required immediately. 

We end by acknowledging that many different groups are working and collaborating to address these issues, especially collision risks. Satellite operators in particular are by no means idle in this regard, as it is in their interest to maintain a safe orbital environment and have teams dedicated to space safety and sustainability. The CRASH Clock is intended to be a tool to help with such efforts. 

\backmatter

\subsection*{Software}
This research was made possible by the open-source projects \texttt{Numba} \cite{numba}, \texttt{Jupyter} \citep{jupyter}, \texttt{iPython} \citep{ipython}, and \texttt{Matplotlib} \citep{matplotlib}.
 
\bibliography{sn-bibliography}{}

\section{Acknowledgments}
We thank Christopher Chyba and Ryne Beeson for helpful discussions regarding our simulation code and the relative velocity calculation. We also thank Hugh Lewis for comments that improved this manuscript.

\section{Author Contributions}
All authors were fully involved in all aspects of this research. ST wrote the initial draft.

FUNDING
Natural Sciences and Engineering Research Council of Canada DH-2022-00477  (AB, SH) and RGPIN-2020-04111 (SL)

\section{Competing interests}
Authors have no competing interests to declare.

\section{Supplementary information}
There is no supplementary material for this work.

\end{document}